\documentclass[fleqn,10pt]{wlscirep}

\usepackage[utf8]{inputenc}
\usepackage{amsfonts}
\usepackage{amssymb}
\usepackage{amsmath}
\usepackage{fullpage}
\usepackage{url}
\usepackage{hyperref}
\usepackage{tikz}
\usepackage{tikz-cd}
\usepackage{graphicx}
\usepackage{subcaption}
\usepackage{underoverlap}
\usepackage{verbatim}
\usepackage{multirow}
\usepackage[normalem]{ulem}

\title{Entropic Analysis of Time Series through Kernel Density Estimation}

\author[1,*]{Audun Myers}
\author[1]{Bill Kay}
\author[1,2]{Iliana Alvarez}
\author[1]{Michael Hughes}
\author[1]{Cameron Mackenzie}
\author[1]{Carlos Ortiz Marrero}
\author[1]{Emily Ellwein}
\author[1]{Erik Lentz}

\affil[1]{Pacific Northwest National Laboratory, Richland, WA, 99354}
\affil[2]{California State University, Los Angeles, CA, 90802}

\affil[*]{audun.myers@pnnl.gov}

\begin{abstract}
    
This work presents a novel framework for time series analysis using entropic measures based on the kernel density estimate (KDE) of a time series' Takens' embedding in two dimensions. Using this framework we introduce two distinct analytical tools: (1) a multi-scale KDE entropy metric, denoted as $\Delta\text{KE}$, which quantifies the evolution of time series complexity across different delay time scales by measuring changes in entropy, and (2) a sliding baseline method that employs the Kullback-Leibler (KL) divergence to detect changes in time series structure through changes in KDEs. 
The $\Delta{\rm KE}$ metric offers insights into the information content and ``unfolding''  properties of the time series' Takens' embedding well-suited to the analysis of dynamical systems, while the KL-divergence-based approach provides a noise and outlier robust measure for identifying time series change points well-suited to the detection of injected radio frequency (RF) signals in noisy backgrounds.  
We demonstrate the effectiveness and versatility of these tools through a set of  experiments across several science and technology domains. In the case of RF communications, we achieve accurate detection of injected signals under varying noise and interference conditions comparable with tools of the art and without the need for training.  Secondly, we apply our methodology to electrocardiography (ECG) data, successfully identifying instances of ventricular fibrillation with high accuracy. Finally, we demonstrate the potential of our tools for identifying anomolous intermittent chaotic regimes within a signal in dynamical systems.

\end{abstract}

\begin{document}
\flushbottom
\maketitle

\section{Introduction}

A \textit{time series} $\mathbf{x} = (x_1, x_2, \ldots, x_t) \in \mathbb{R}^t$ is a real-valued vector where $\{x_i\}_{i=1}^t$ typically represent data observations sequentially sampled at equal time intervals from some sensor or system. Time series data is ubiquitous across scientific domains---in this document alone we draw examples from  radio frequency (RF) signal analysis~\cite{o2018over}, biomedical data~\cite{nolle1986crei}~\cite{goldberger2000physiobank}, and dynamical systems~\cite{pomeau1980intermittent}; there are many other examples including seismology, finance, and meteorology.



In dynamical systems research, there is a pervasive goal of classifying when and to what extent time series data is chaotic. 
In~\cite{fraser1986independent} it was observed that subsampling of time series from dynamical systems at different time scales resulted in a corresponding ``unfolding of the attractor'' instinctive phenomenon which could be readily discovered when a system is periodic, partially observed if the system is quasi-periodic, and not at all observed if the system is chaotic.

Motivated by the unfolding of the attractor phenomenon, we present novel quantitative methods herein which measure information-theoretic changes in time series. To cast time series analysis in the regime of information theory, we first apply a \textit{Takens' embedding} to map the time series to a point cloud. We then approximate the point cloud with a probability density function (PDF) via kernel density estimation. Finally, we compute the entropy of the approximate PDF. This entropy value, and its change over time, contains valuable information about the degree to which the time series is unpredictable. In many settings, this can be interpreted as a measure of the amount of noise present in the time series; as a proxy to the \textit{signal-to-noise ratio} (SNR) of RF signals is one such example.

The remainder of the document is layed out as follows:  In Section~\ref{sec:tbg} we introduce each piece of our KDE-entropy computations individually. In Section~\ref{sec:methods} we describe how to join these pieces together into a workflow with some motivating examples from RF analysis. In Section~\ref{sec:results} we present three domain applications motivating the effectiveness of our methods. We include results from RF analysis, cardiology, and dynamical systems. In Section~\ref{sec:conc} we close with some conclusions and future directions. 

\section{Technical Background}
\label{sec:tbg}

In this section, we introduce each component of our workflow as a self-contained analytical tool, with Section~\ref{sec:te} describing the Takens' embedding, Section~\ref{sec:kde} describing the kernel density estimate (KDE), and Section~\ref{sec:ekl} describing entropy and the KL divergence of distributions. How these components will be combined into workflows will be covered in Section~\ref{sec:methods}.   


\subsection{Takens' Embeddings}
\label{sec:te}

\begin{figure}[h!]
\begin{center}
    \includegraphics[width=0.99\textwidth]{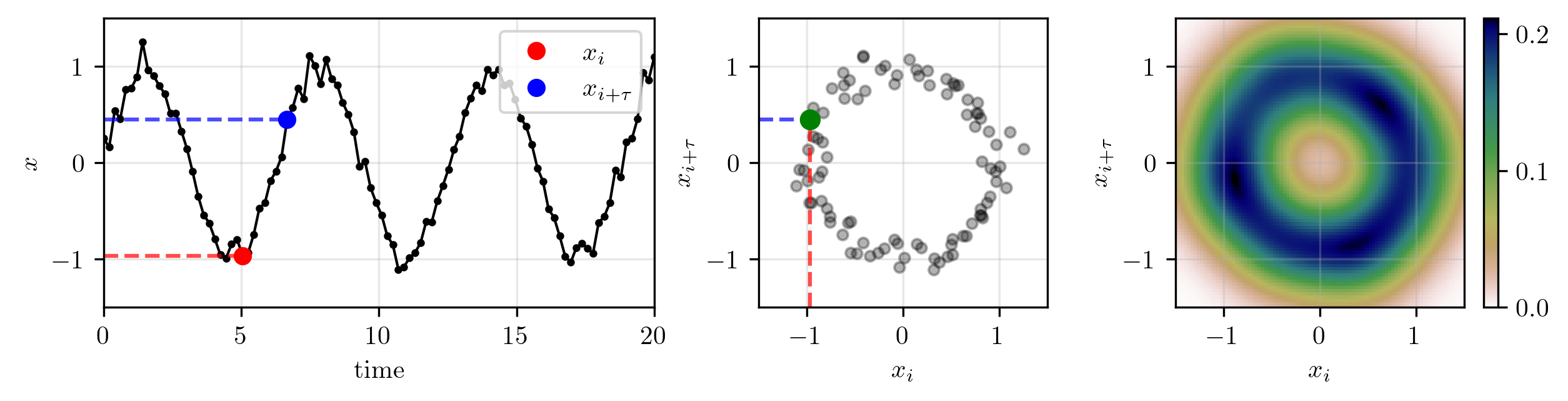}
    \caption{Sinusoidal time series with additive Gaussian noise (left), two-dimensional Takens' embedding of the time series (center), and resulting kernel density estimation of the Takens' embedding (right).}
    \label{fig:takens_sine}
\end{center}
\end{figure}

Given a time series $\mathbf{x}=(x_1, x_2, \ldots, x_t) \in \mathbb{R}^t$, 
the ($2$-dimensional) Takens' embedding is given by
\begin{equation}
    P_{\tau}(\mathbf{x}) := \{(x_i, x_{i+\tau})\}_{i = 1}^{t - N}\text{ for }1 \leq \tau \leq N
    \label{eq:takens}
\end{equation}
where $\tau, N \in \mathbb{N}$ are a time delay parameter and a delay upper bound. the left-most two panels of Figure~\ref{fig:takens_sine} shows how a sinusoidal wave with additive Gaussian noise is transformed into a point cloud via the Takens' embedding. 

\subsection{Kernel Density Estimation}
\label{sec:kde}

%


KDE is a method of approximating an underlying probability density function from a given set of data~\cite{chen2017tutorial}. Intuitively, one can think of KDE as a complementary procedure to sampling, where in sampling the input is a PDF and the output is a collection of data points which approximate that PDF. In KDE, the input is a collection of data points and the output is a PDF which approximates those data points. In~\ref{sec:appendkde} we provide details about how KDE is often accomplished, including the parameter selections used in this work.  In this document we use a Gaussian kernel, meaning the PDFs produced by KDE are sums of Gaussian distributions and are thus continuous PDFs regardless of the nature of the input data. Parameters not specified in ~\ref{sec:appendkde} use the default parameters from the \texttt{gaussian\_kde} method in the \texttt{scipy.stats} package~\cite{2020SciPy-NMeth}. The two right-most panels in Figure~\ref{fig:takens_sine} illustrate a simple example of the KDE of the ciruclar point cloud from the sinusoidal wave as a simple example where the height or probability density is represented with a heatmap.

\subsection{Entropy and KL Divergence}
\label{sec:ekl}
Given any (continuous) random variable $X$ with  support $\mathcal{X}$ and PDF $p(\cdot)$, the (differential) Shannon entropy of $X$ is given by
\begin{equation}
H(X):= -\int_{\mathcal{X}} p(x) \log_2p(x) \, \mathrm{d}x ,
\label{eq:shannon_entropy}
\end{equation}
and expresses the expected number of bits one needs to convey the outcome of $X$. 

Closely related to the Shannon entropy of a PDF is the \textit{Kullback-Leibler} (KL) divergence~\cite{kullback1951information}  between two PDFs $p$ and $q$ with joint support $\mathcal{X}$, which is defined as
\begin{equation}
D(p || q) := \int_{ \mathcal{X}} p(x) \log_2\left (\frac{p(x)}{q(x)}\right)\, \mathrm{d}x ,
\label{eq:cont_KL}
\end{equation}
The KL divergence is a measure of how different two PDFs are. While not a metric (failing symmetry and the triangle inequality), the KL divergence is an industry standard proxy for the distance between two distributions~\footnote{One reason for this is that the KL divergence is the expected log-likelihood ratio between two distributions~\cite{kullback1951information}. For an overview of KL divergence, see~\cite{cover1991information}.}. There are analogous versions of Equations~\ref{eq:shannon_entropy} and~\ref{eq:cont_KL} for discrete  random variables~\cite{cover1991information}, and while many of the applications herein we are indeed working with continuous PDFs, the entropy calculations are all evaluated by numerical integration methods, and so the discrete versions are the ones we deploy in practice.  
\section{Methods}
\label{sec:methods}

In this section, we describe how to combine the tools from Section~\ref{sec:tbg} into a coherent workflow for time series analysis. Our approach uses entropy measures computed from KDEs of Takens' embeddings across multiple time scales to quantify signal complexity and detect changes in time series dynamics. We present two complementary methods: (1) \textit{Kernel Density Estimate Entropy} (KDEE), which measures signal complexity by analyzing entropy changes across time scales, and (2) \textit{Sliding Baseline Change Detection} (SBCD), which uses KL divergence between KDEs to identify structural changes in signals. 


Our methodology follows a consistent pipeline regardless of the specific application:

\begin{enumerate}
    \item \textbf{Embedding}: Apply Takens' embedding (Equation~\ref{eq:takens}) to transform the time series $\mathbf{x}$ into point clouds $P_\tau(\mathbf{x})$ for various time delays $\tau$
    \item \textbf{Density Estimation}: Compute KDE $K(P_\tau(\mathbf{x}))$ for each embedded point cloud using a Gaussian kernel
    \item \textbf{Entropy Calculation}: Calculate Shannon entropy $H(K(P_\tau(\mathbf{x})))$ of each KDE
    \item \textbf{Analysis}: Apply either complexity analysis or change detection based on the entropy measures
\end{enumerate}

\noindent Figure~\ref{fig:takens_sine} illustrates the KDE of a Takens' embedding of a noisy sinusoid, which constitutes steps (1)-(3) of this pipleline.


For computational efficiency and to avoid the curse of dimensionality, we use 2-dimensional embeddings ($m=2$) throughout. While optimal embedding methods exist~\cite{kennel1992determining,fraser1986independent}, our experiments show that $m=2$ effectively captures sequential relationships for complexity analysis while enabling efficient computation and clear visualization.

\subsection{Kernel Density Estimate Entropy (KDEE)}
\label{ssec:kdee}

For a given time delay $\tau$, we define the \textit{Kernel Density Estimate Entropy} (KDEE) as the composite process
\begin{equation}
    \text{KE}_\tau(\mathbf{x}) := H(K(P_\tau(\mathbf{x}))).
\end{equation}
While $\text{KE}_\tau(\mathbf{x})$ provides information about signal structure at time scale $\tau$, it is sensitive to sampling frequency and other signal properties independent of the underlying dynamics. To address this limitation, we analyze the evolution of KDEE across multiple time scales.

\begin{figure}[h!]
\begin{subfigure}{.49\textwidth}
\centering
\includegraphics[width=.99\linewidth]{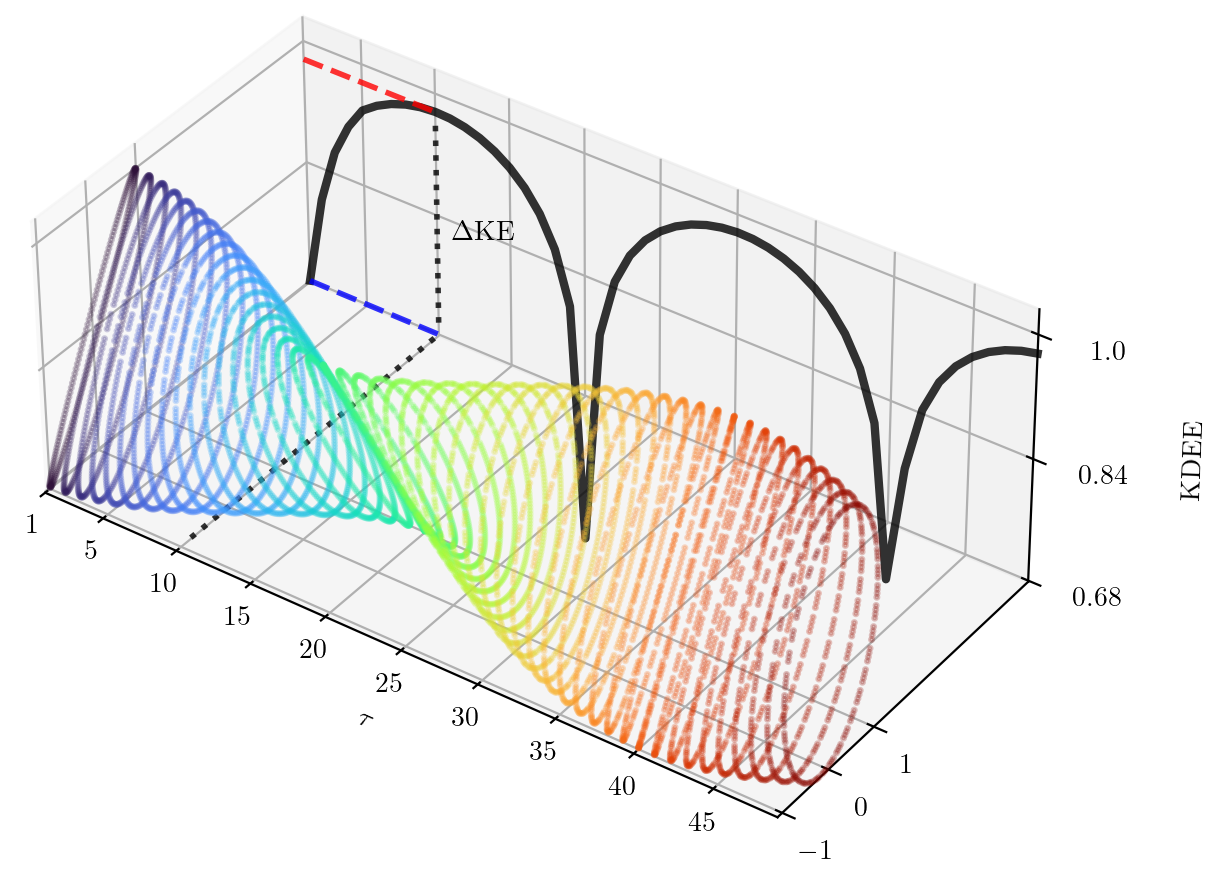}
\caption{Sinusoidal signal showing attractor unfolding}
\end{subfigure}
\hfill
\begin{subfigure}{.49\textwidth}
\centering
\includegraphics[width=.99\linewidth]{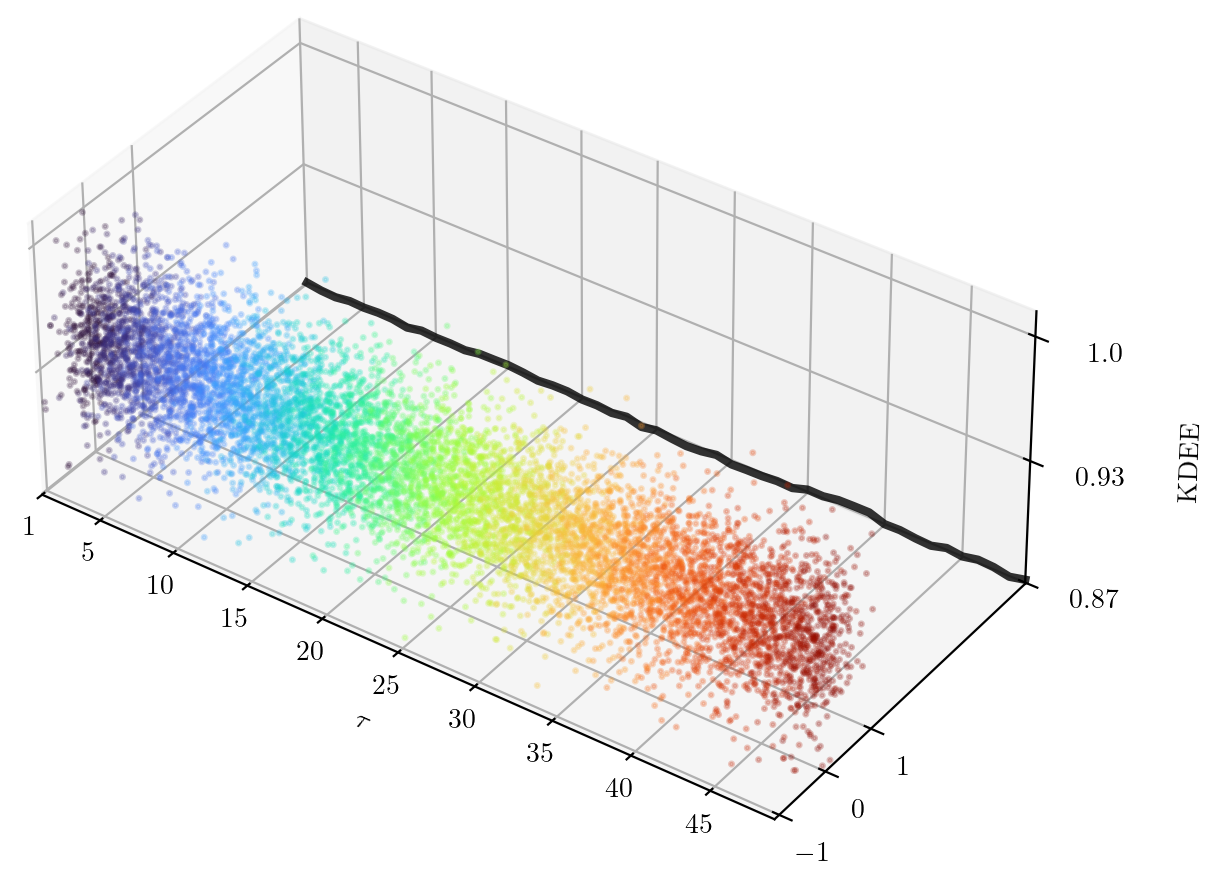}
\caption{Gaussian noise showing minimal variation}
\end{subfigure}
\caption{KDEE evolution across time scales $\tau$ for structured signal (left) versus noise (right), showing characteristic unfolding behavior.}
\label{fig:multi-scale_KDE}
\end{figure}

The key insight is that different signal types exhibit characteristic patterns as $\tau$ varies. Structured signals (e.g., periodic) show significant entropy variation as their attractors ``unfold'' with increasing $\tau$, while noise-dominated signals show minimal variation. We quantify this behavior using
\begin{equation}
\Delta\text{KE}(\mathbf{x}) = \max_{\tau}\text{KE}_\tau(\mathbf{x}) - \min_{\tau'}\text{KE}_{\tau'}(\mathbf{x})
\end{equation}
Figure~\ref{fig:multi-scale_KDE} demonstrates this concept, comparing a sinusoidal signal with pure Gaussian noise. The structured signal shows clear entropy variation as the embedding unfolds, while noise exhibits minimal change. Further, in practice, $\Delta\text{KE}(\mathbf{x})$ serves as a proxy for signal-to-noise ratio, with larger values indicating more structured (less noisy) signals.

\subsection{Sliding Baseline Change Detection}
\label{ssec:klkde}

To detect changes in signal dynamics over time, we use a sliding baseline approach that compares the KDE of a current window against a robust baseline derived from previous windows.

\textbf{Baseline Construction}: For each analysis window, we compute KDEs from the previous $W$ windows and take their pointwise median to create a robust baseline $K_{\text{baseline}}$. The median provides robustness against outliers, requiring only that $>50\%$ of baseline windows contain uncontaminated data.

\textbf{Change Detection}: We measure the dissimilarity between the baseline and current window using a symmetrized KL divergence:
\begin{equation}
    \bar{D}_{KL}(P,Q) = \frac{1}{2}(D_{KL}(P||Q) + D_{KL}(Q||P)).
\end{equation}
To handle numerical issues with zero probabilities, we add regularization terms $0.001 \max(P)$ and $0.001 \max(Q)$ to both distributions before computing the divergence.

\begin{figure}[h!]
\begin{center}
    \includegraphics[width=0.95\textwidth]{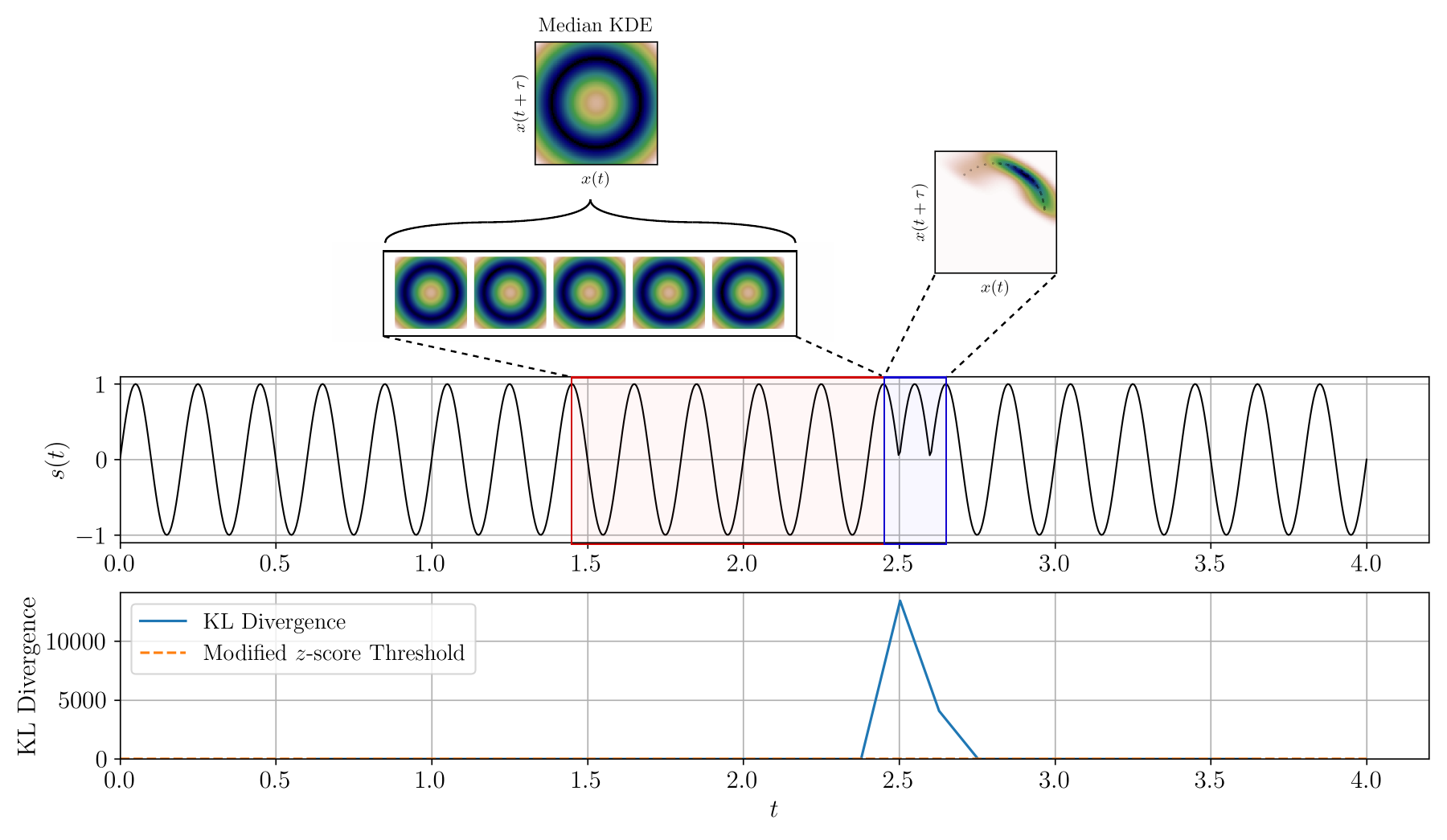}
    \caption{Change detection using KL divergence between median baseline (red windows) and analysis window (blue). Bottom panel shows detection statistic with threshold.}
    \label{fig:methods_kde_kl}
\end{center}
\end{figure}

\textbf{Significance Testing}: We identify significant changes using the modified $z$-score
\begin{equation}
Z = \frac{0.6745(X - \text{median}(X))}{\text{MAD}},
\end{equation}
where MAD is the median absolute deviation. Values with $|Z| > 3.5$ indicate significant deviations from baseline behavior. 
Figure~\ref{fig:methods_kde_kl} illustrates this approach on a signal containing a brief structural change from $\sin(2\pi t)$ to $|\sin(2\pi t)|$. This approach enables robust detection of structural changes in signal dynamics while maintaining computational efficiency and providing interpretable results through entropy-based measures.

\section{Results}
\label{sec:results}

In this section, we evaluate the effectiveness of our KDE-based entropic signal measures across diverse scientific domains to demonstrate their broad applicability and robustness. We apply our methodology to three distinct areas: RF signal processing for injection detection (Section~\ref{sec:RF}), ECG analysis for ventricular fibrillation detection (Section~\ref{sec:ECG}), and dynamical systems analysis for state detection in chaotic systems (Section~\ref{sec:dynamic}).

To provide rigorous performance evaluation, we benchmark our methods against established state-of-the-art techniques. For unsupervised anomaly and change point detection in time series, autoencoder (AE)-based neural networks represent a leading methodology~\cite{malhotra2015long, an2015variational, zamanzadeh2024deep}. These models excel at learning robust, low-dimensional representations of normal temporal patterns from training data, with anomalies identified through significantly higher reconstruction errors that indicate deviation from learned baseline patterns. Their ability to capture complex nonlinear dependencies without requiring labeled anomalous training data makes them an appropriate benchmark for assessing our method's performance across multiple application domains. Implementation details for our autoencoder comparison are provided in~\ref{app:AE}.

\subsection{Radio Frequency Signal Processing for Injection Detection}
\label{sec:RF}

Radio frequency signals, spanning the frequency spectrum from 30 kHz to 300 GHz~\cite{scarpati2021radio} and propagating within/using a variety of media (gases, liquids, solids, electromagnetic fields, etc.), are commonly recorded as time series for analysis. This investigation focuses on detecting informative engineered communication signals superposed with noisy environments, independent of the specific modulation scheme employed and prior to demodulation (over-the-air signals~\cite{o2018over}). This approach enhances the generalizability of our methods across various modulation techniques, a consideration of practical importance in RF signal analysis.

While entropy-based methods have been explored for classifying received signals by modulation type~\cite{kay2024permutationentropysignalanalysis}, we concentrate on the fundamental problem of signal identification in the presence of noise and interference. The primary objective is to detect the injection of structured signals into noisy environments and identify their temporal locations.

\textbf{Signal Generation and Experimental Setup}

We employ simulated signals generated using the framework from~\cite{o2018over}, which provides precise control over Signal-to-Noise Ratio (SNR) across 14 modulation formats\footnote{Supported modulation formats: BPSK, QPSK, OQPSK, Pi4QPSK, 8PSK, 16PSK, OOK, 4ASK, 8ASK, 16QAM, 64QAM, 32QAM, 16APSK, and 32APSK.}. To create realistic conditions, we additionally simulate background interference representing narrow frequency bands of competing signals as would be encountered in congested RF environments. Details of the interference simulation are provided in~\ref{app:SIR}.

For our experiments, ten signals were generated for each modulation type across all background simulations. Each signal comprised 5000 samples representing 100 symbols (50 samples per symbol and 1 cycle per symbol) over a 1-second interval at unit root-mean-squared amplitude. This results in a sampling rate of $5000 \text{ Hz}$ ($F_s = 5000 \text{ Hz}$) and signal frequency of 100 Hz, which is used consistently throughout the simulation. A raised-cosine filter with roll-off factor 0.25 was applied. Both additive Gaussian noise and interference were introduced to achieve target SNR and Signal-to-Interference Ratio (SIR) values, with SNR calculated using the ratio of signal and noise power estimated from root-mean-square values (detailed in~\ref{app:SNR}).

A randomly selected time window within the latter half of each signal was designated as the injection interval, with duration randomly chosen between 20\% and 40\% of the total signal length. During injection periods, the signal with specified SNR, SIR, and modulation was superimposed onto the background. Outside these periods, only noise and simulated interference signals were present.

\textbf{Theoretical Motivation for Detector Usage: Correlation Between $\Delta\text{KE}$ and SNR}

\begin{figure}[h!]
\begin{center}
    \includegraphics[width=0.95\textwidth]{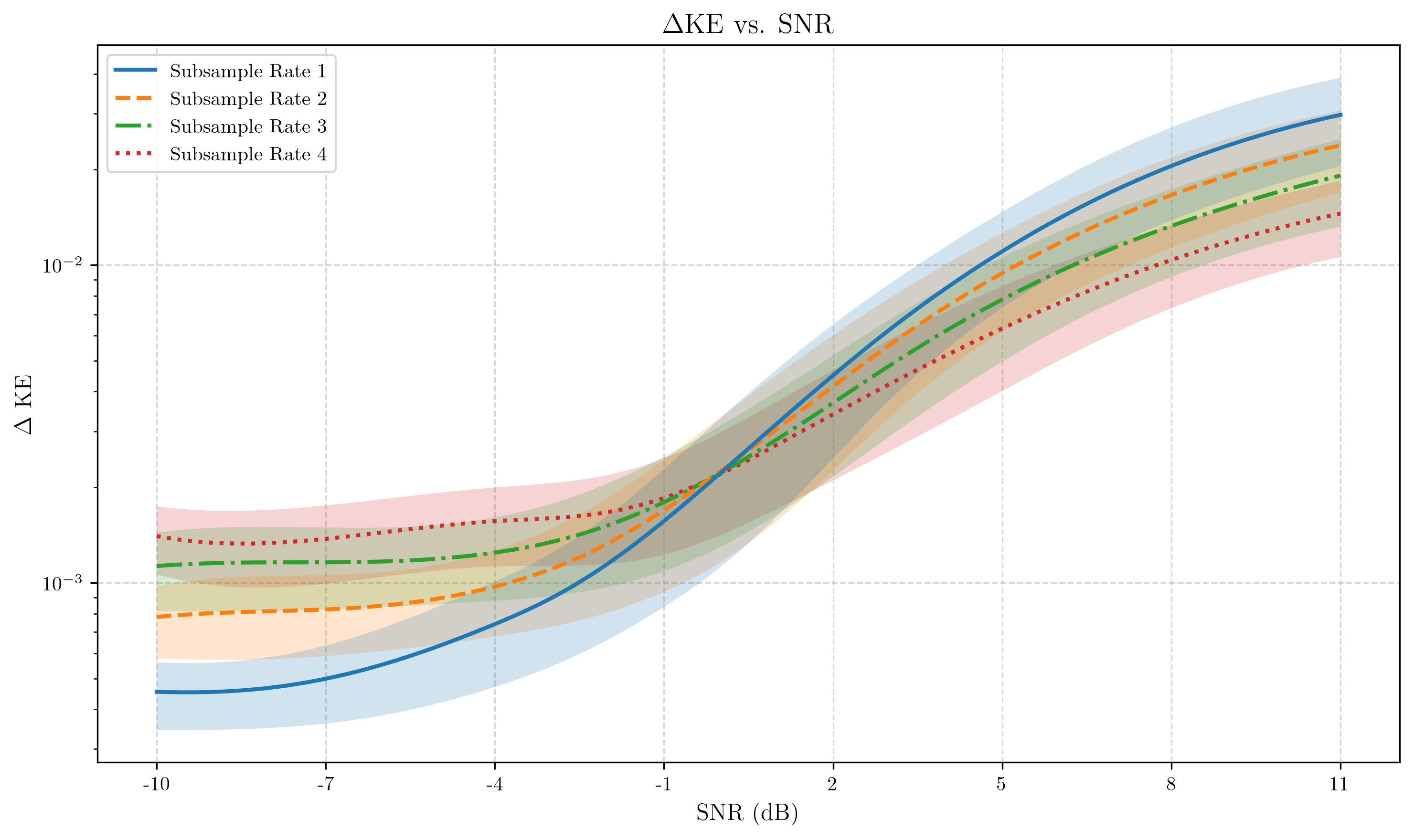}
    \caption{Correlation between SNR and $\Delta\text{KE}$ across modulation types. For each SNR, 10 signals of each of 14 modulation types were generated. Results show means with one standard deviation on a log scale, demonstrating consistent correlation independent of modulation scheme.}
    \label{fig:delta_ke}
\end{center}
\end{figure}

To provide theoretical foundation for using $\Delta\text{KE}$ in RF analysis, we establish its correlation with SNR. In practice, the SNR of observed signals is unknown, making $\Delta\text{KE}$ valuable as a directly observable proxy. We demonstrate that $\Delta\text{KE}$ correlates with SNR across various sampling rates and serves as a reliable SNR indicator in noisy regimes (down to approximately -10 dB). Using the signal simulation framework, we generated a corpus of 10 signals for each modulation type at SNRs from -10 to 11 dB in 3 dB steps, computing $\Delta\text{KE}$ values for signals of length 3000 samples, downsampled by factors of 2, 3, and 4. Figure~\ref{fig:delta_ke} shows the mean and standard deviation of results averaged across modulation types.

The results show minimal variation in the SNR-$\Delta\text{KE}$ correlation across individual modulation types, indicating that this metric provides a useful proxy for signal structure assessment when modulation type is unknown. This correlation motivates using $\Delta\text{KE}$ as a change-point detector, since changes in $\Delta\text{KE}$ reliably indicate changes in signal structure.

\textbf{Detection Performance Analysis}

\begin{figure}[h!]
\begin{center}
    \includegraphics[width=0.95\textwidth]{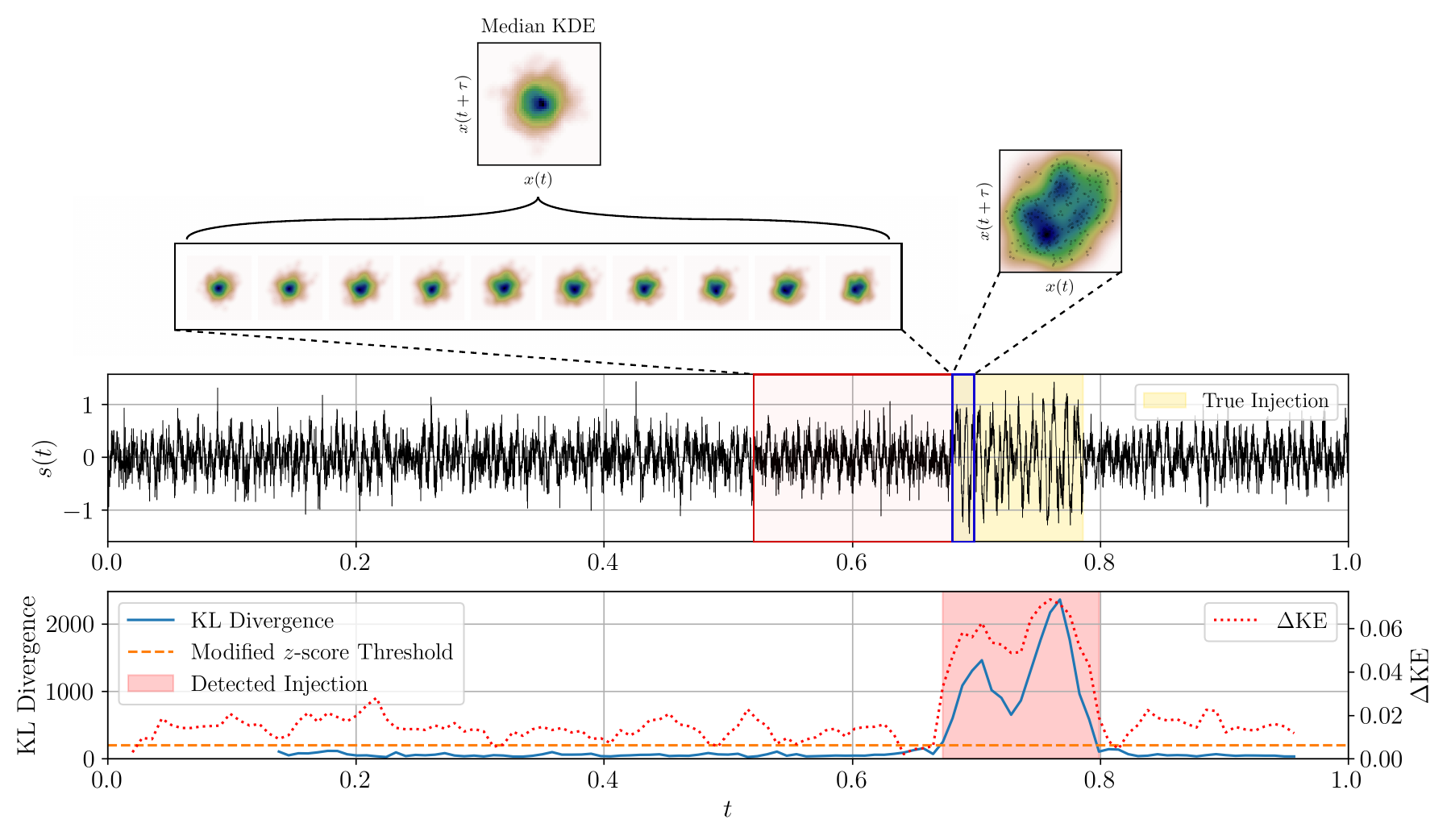}
    \caption{RF signal injection detection example showing background interference, noise, sliding baseline (red window, 10 KDEs), and analysis window (blue window). Both KL divergence and $\Delta\text{KE}$ methods show clear response during injection period, with detected region (red) based on modified z-score threshold.}
    \label{fig:RF_example_sliding_baseline}
\end{center}
\end{figure}

Figure~\ref{fig:RF_example_sliding_baseline} demonstrates our methodology applied to a sample signal with SIR = SNR = 4 dB and BPSK modulation. The figure shows the sliding baseline construction using 10 KDEs with their median, the analysis window during signal transition, and detection results from both KL divergence and $\Delta\text{KE}$ methods.

\begin{figure}[h!]
\begin{center}
    \includegraphics[width=0.99\textwidth]{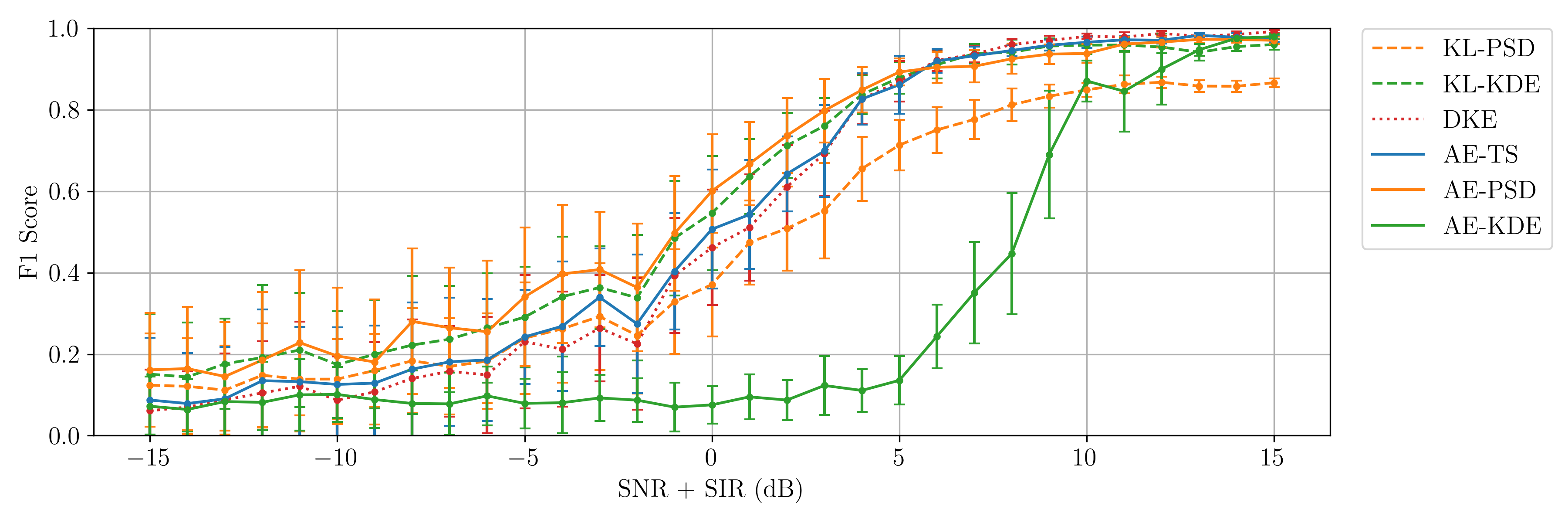}
    \caption{RF signal injection detection performance comparison across varying SNR and SIR conditions. Methods evaluated using raw Time Series (TS), Power Spectral Density (PSD), and Kernel Density Estimate (KDE) representations with Kullback-Leibler divergence (KL) and Autoencoder (AE) detection approaches.}
    \label{fig:RF_detection_results}
\end{center}
\end{figure}

We conducted comprehensive comparative analysis across three detection methodologies ($\Delta\text{KE}$, KL divergence, and autoencoder-based approaches) and three data representations (raw time series, power spectral density, and KDE from Takens' embedding). Performance was quantified using F1 scores, calculated by detecting windows with at least 25\% overlap with ground truth injection intervals. Figure~\ref{fig:RF_detection_results} presents F1 scores with standard deviation across a range of SNR+SIR, where the SIR and SNR are the same value and the amplitude of their signal's is calculated separately and then added. This results in the total background (noise and interference) having a signal to background ratio lower than the individual SIR and SNR value. Results were derived from ten independent trials of each modulation type and background signal combination. Minimal performance variation across modulation types allowed integration of all trials into single F1 scores per SNR+SIR level, further demonstrating modulation-independent robustness.

Among the top-performing methods (KL-KDE, AE-TS, and AE-PSD), the KL-KDE approach uniquely requires no neural network training while achieving comparable performance to autoencoder-based methods. This computational advantage makes KL-KDE particularly suitable for streaming data applications requiring real-time change detection. We conjecture that AE-KDE's particularly poor performance was due to its robustness; the reconstruction loss was low during the injected time period, causing false negative responses. 

\begin{table}[h!]
\caption{Mean computation time for single RF signal with sliding windows.}
\centering
\begin{tabular}{|cc|c|}
\hline
\multicolumn{1}{|c|}{\textbf{Detection Method}} & \textbf{Representation} & \textbf{Mean Compute Time (s)} \\ \hline
\multicolumn{1}{|c|}{\multirow{3}{*}{Autoencoder}} & Time Series & 0.441 \\ \cline{2-3}
\multicolumn{1}{|c|}{} & Power Spectral Density & 0.387 \\ \cline{2-3}
\multicolumn{1}{|c|}{} & Kernel Density Estimate & 0.523 \\ \hline
\multicolumn{1}{|c|}{\multirow{2}{*}{KL Divergence}} & Power Spectral Density & 0.018 \\ \cline{2-3}
\multicolumn{1}{|c|}{} & Kernel Density Estimate & 0.066 \\ \hline
\multicolumn{2}{|c|}{$\Delta\text{KE}$} & 0.398 \\ \hline
\end{tabular}
\label{tab:RF_compute_time}
\end{table}

Table~\ref{tab:RF_compute_time} reveals significant computational advantages for KL divergence-based methods compared to autoencoder approaches. This efficiency stems from the inherent complexity of autoencoder training and inference, which involves substantial matrix operations and iterative optimization, whereas KL divergence calculations are comparatively lightweight. While $\Delta\text{KE}$ exhibits processing speeds similar to autoencoders, its detection performance falls short of the KL-KDE approach, highlighting the computational benefits of KL divergence methods for applications requiring rapid analysis capabilities.

\subsection{ECG Data Analysis}
\label{sec:ECG}

Electrocardiogram data provides time series representations of cardiac electrical activity commonly used for diagnosing various medical conditions~\cite{hong2019mina,zhang2020anomaly}. Anomalies in ECG readings can correspond to cardiac events of clinical interest, making ECG analysis an important application domain for time series anomaly detection methods. While various approaches (both machine learning and domain knowledge-based) have been applied to ECG anomaly detection, we emphasize that this work makes no claims regarding clinical efficacy. The following is intended only as an exercise. Development of entropic KDE measures for diagnostic purposes would require future collaboration with appropriate clinical professionals.

We evaluate our entropic KDE statistics on the PhysioNet CU Ventricular Tachyarrhythmia Database~\cite{nolle1986crei,goldberger2000physiobank}, analyzing 35 eight-minute ECG recordings from subjects who experienced episodes of sustained ventricular tachycardia, ventricular flutter, and ventricular fibrillation (VF). Each signal includes clinical state labels, with all abnormal ventricular-related states classified as VF for binary anomaly detection (normal versus abnormal).

\begin{figure}[h!]
\begin{center}
    \includegraphics[width=0.95\textwidth]{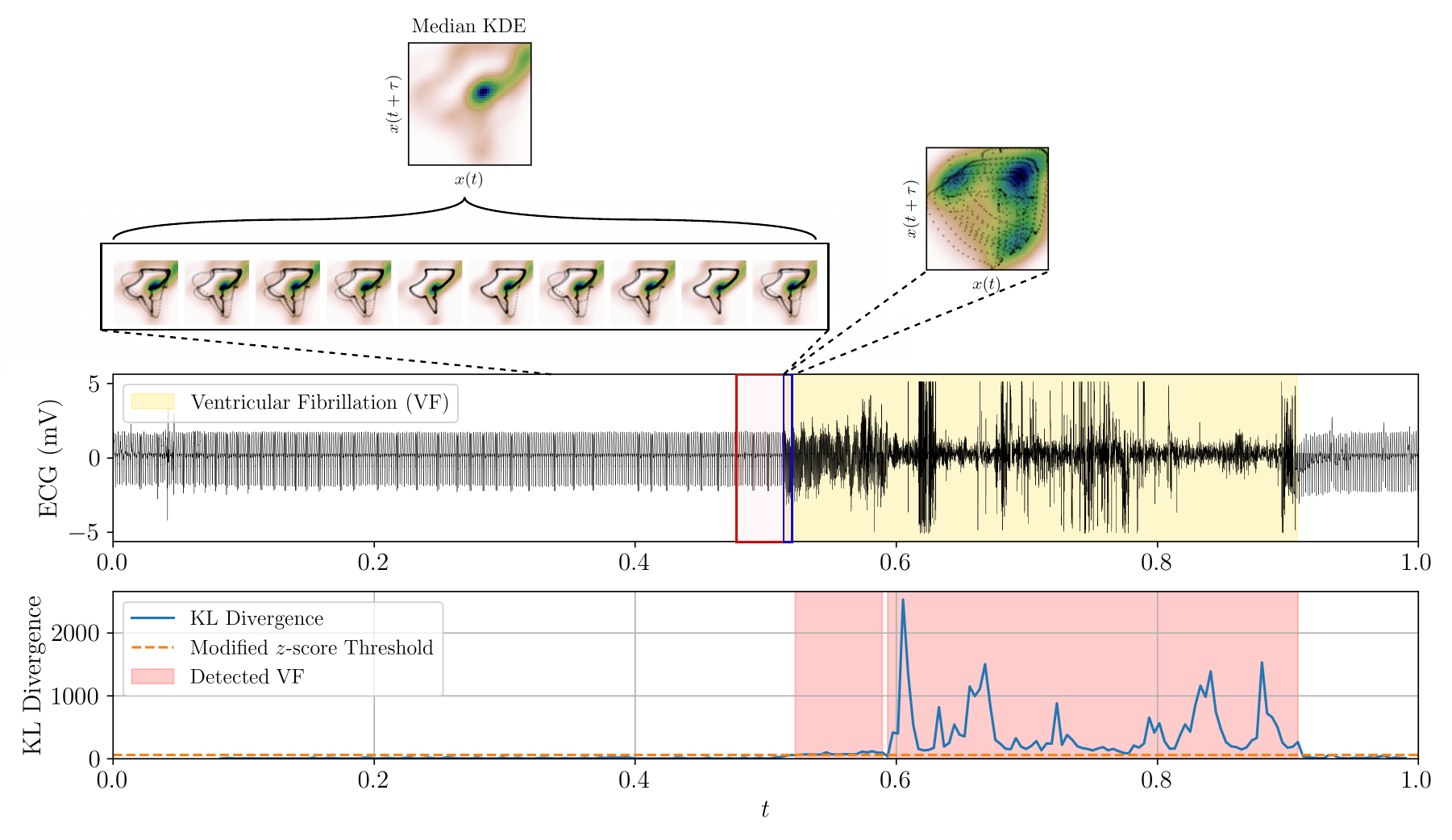}
    \caption{ECG signal analysis example showing sliding baseline construction with median KDE compared to analysis window. Detected VF interval (red) demonstrates strong alignment with clinically labeled VF period (yellow).}
    \label{fig:ECG_example_sliding_baseline}
\end{center}
\end{figure}

Figure~\ref{fig:ECG_example_sliding_baseline} illustrates our methodology applied to an ECG signal from the dataset. The baseline windows (highlighted in red) and analysis window (blue) are used to compute KL divergence between the current window's KDE and the median baseline KDE. The resulting KL divergence values are shown with detected VF periods identified where values exceed the modified z-score threshold, demonstrating strong alignment between detected and clinically labeled intervals.

\begin{table}[h!]
\caption{Ventricular fibrillation detection performance across methods and representations.}
\centering
\begin{tabular}{|c|c|c|}
\hline
\textbf{Detection Method} & \textbf{Representation} & \textbf{F1 Score} \\ \hline
\multirow{3}{*}{Autoencoder} & Time Series & $80.3\% \pm 12.1\%$ \\ \cline{2-3}
& Power Spectral Density & $81.2\% \pm 13.1\%$ \\ \cline{2-3}
& Kernel Density Estimate & $62.5\% \pm 16.3\%$ \\ \hline
\multirow{2}{*}{KL Divergence} & Power Spectral Density & $77.8\% \pm 10.2\%$ \\ \cline{2-3}
& Kernel Density Estimate & \textbf{$87.1\% \pm 8.7\%$} \\ \hline
\end{tabular}
\label{tab:ecg_tab_results}
\end{table}

Table~\ref{tab:ecg_tab_results} presents F1 scores across all tested methods and representations. The KL divergence approach applied to KDE representations of Takens' embeddings achieved the highest performance (87.1\% ± 8.7\%), outperforming all autoencoder-based methods while maintaining the computational efficiency advantages demonstrated in the RF analysis. This superior performance suggests that the geometric structure captured by Takens' embeddings and quantified through KDE entropy measures provides particularly effective features for detecting physiological state changes in cardiac rhythms. We hypothesize that the AE did not perform as well when using KDE due to it being overly sensitive to slight changes in the distribution in comparison to the KL divergence.

\subsection{Dynamic State Detection}
\label{sec:dynamic}

To demonstrate the broader applicability of our entropic KDE framework beyond engineered signals, we now turn our methods to detecting periodic and chaotic regimes within intermittent dynamical systems characterized by irregular transitions between ordered and disordered behavior~\cite{pomeau1980intermittent}. We analyze the $x$ variable of the Lorenz system:
\begin{equation}
\begin{aligned}
\frac{dx}{dt} &= \sigma(y - x), \\
\frac{dy}{dt} &= x(\rho - z) - y, \\
\frac{dz}{dt} &= xy - \beta z,
\end{aligned}
\label{eq:lorenz}
\end{equation}
using standard parameters $\sigma = 10$, $\beta = 8/3$, and $\rho = 166.18$ that exhibit type-I intermittency. The system was numerically integrated at 150 Hz for 1000 seconds using the \texttt{teaspoon} Python package~\cite{myers2020teaspoon}, with the initial 93 seconds discarded to ensure fully developed dynamics.

\begin{figure}[h!]
\begin{center}
    \includegraphics[width=0.95\textwidth]{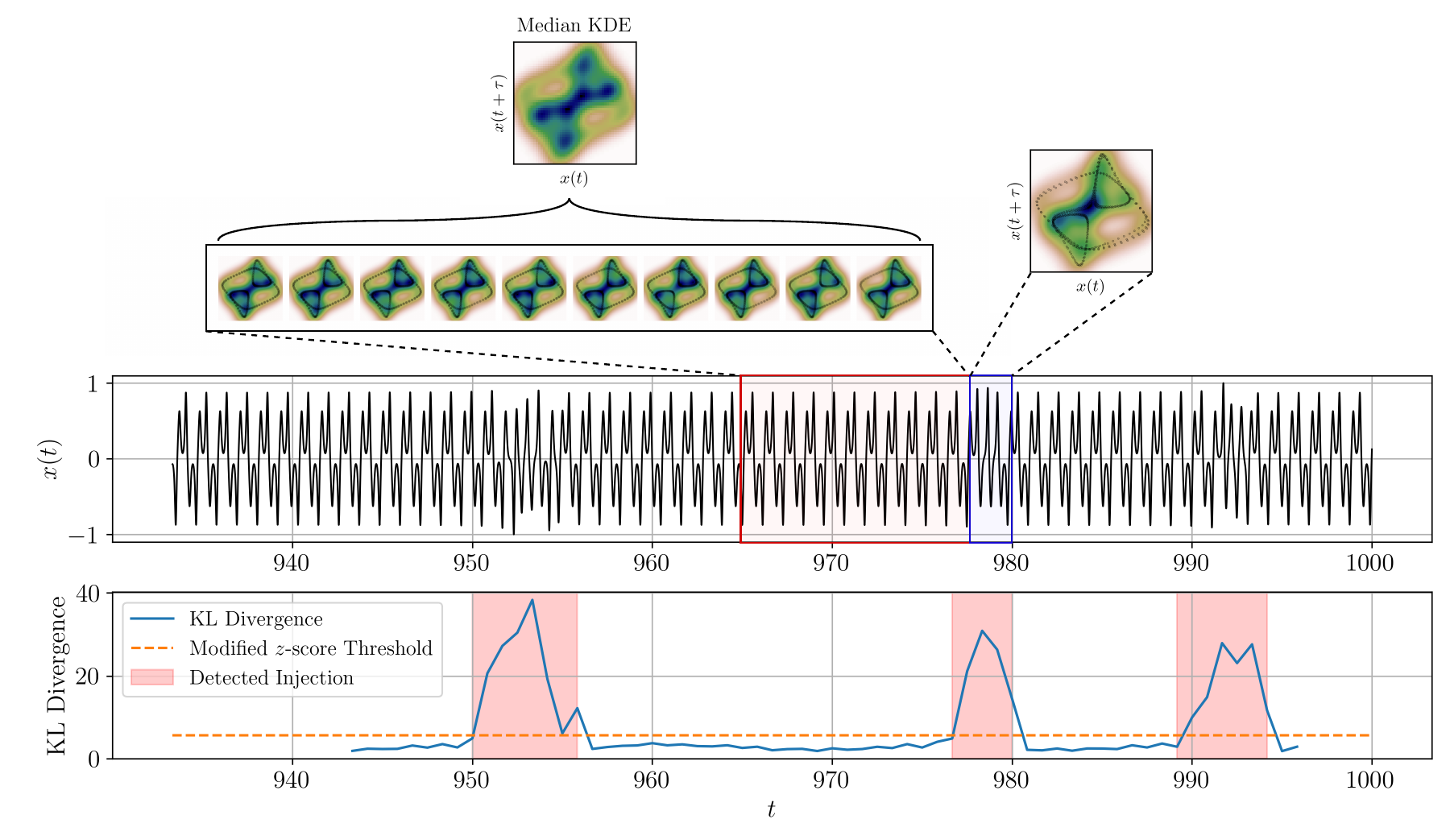}
    \caption{Dynamical systems analysis of intermittent Lorenz system showing detection of chaotic intervals using sliding baseline approach. KL divergence peaks correspond to transitions between periodic and chaotic regimes, demonstrating the method's sensitivity to fundamental changes in system dynamics.}
    \label{fig:INT_example_sliding_baseline}
\end{center}
\end{figure}

Figure~\ref{fig:INT_example_sliding_baseline} illustrates the time series $x(t)$ and corresponding chaotic interval detection results. The KL divergence approach successfully identifies chaotic intervals through peaks in the divergence measure, as shown in the bottom panel. These results demonstrate that our entropic KDE framework can detect fundamental changes in dynamical system behavior, opening avenues for future research in applying KDE-based methods to the study of complex dynamical systems.

\section{Conclusion}
\label{sec:conc}
In this document we introduced a novel analytical composite pipeline for analyzing qualitative properties of time series by transforming them into point clouds using two-dimensional Takens' embeddings, then approximating those point clouds with a PDF using KDE, and lastly computing the PDF's entropy. 
The resulting tool, KDEE, was then used on the time series of three distinct data driven use-cases- RF signal processing, cardiology, and dynamical systems- to demonstrate proof-of-concept utility across a broad set of applications. 
The demonstrations showed robust identification of the following qualitative and quantitative properties in each domains' time series: (1) detecting the transient injection of RF communications signals in a noisy background at competitive sensitivity to state-of-the-art AE algorithms using raw time series or PSD as input; (2) classification of VF episodes in ECG data using a simple KL-divergence threshold with higher accuracy than AE anomaly detectors trained on the labeled data; (3) identification of chaotic periods in non-linear dynamical systems. Further, these results were attainable without extensive training regiments and a far smaller computaitonal footprint than their learning counterparts. The success across these three diverse domains provides a first sampling of what we beilieve is a broad applicability and robustness of our entropic KDE framework for detecting structural changes in time series data across technical disciplines.

\section{Acknowledgements}
This work is funded by the Pacific Northwest National Laboratory.
Pacific Northwest National Laboratory information release number PNNL-SA-209698.

\bibliography{bib}

\appendix{}

\section{Kernel Density Estimation (KDE)}
\label{sec:appendkde}

In this appendix, we provide relevant background on kernel density estimation towards keeping the methods of this document self-contained. For a thorough treatment of kernel density estimates, see ~\cite{chen2017tutorial}.

Developed independently by Parzen and Rozenblatt~\cite{parzen1962estimation}\cite{davis2011remarks} kernel density estimation provides a partial answer to the following question:\textit{ Given $(x_1, x_2, \ldots, x_n)$ a set of samples from $\mathbb{R}^d$ drawn independently and identically distributed from an unknown PDF $f$, how can we estimate $f$?} A partial answer to this question is provided by the (family of) \textit{kernel density estimates} with \textit{kernel} $K$ and smoothing \textit{bandwidth} parameter $h$ (denoted $\hat{f}_{h, K}$), which are formally defined as:

\[
\hat{f}_{h,K}(x):= \frac{1}{nh^d}\sum_{i=1}^n K\left(\frac{x-x_i}{h}\right)
\]
where $K: \mathbb{R}^d \rightarrow \mathbb{R}$ is a smooth function called a \textit{kernel} and $h > 0$ is called a bandwidth parameter.  Selection of $K$ and $h$ is itself an active research area, but absent external domain knowledge steering a decision, a common choice (and the one made in this document) for $K$ is the \textit{Gaussian kernel} $\phi(x)$, given by the PDF for the standard normal distribution:
\[
\phi(x)= \frac{1}{\sqrt{2 \pi}} \texttt{exp}\left\{\left (-x^2/2\right)\right\}.
\]
Selecting $h$ is a difficult problem, with entire works dedicated to the survey of bandwidth selection (see, e.g.,~\cite{park1990comparison}). Choosing a bandwidth is a balance between under and over smoothing; an optimal choice of bandwidth typically seeks to minimize the \textit{mean integrated suare error}:

\[
\text{MISE}(h):= \mathbb{E}\left\{\int \left (\hat{f}_{h,K}(x)-f(x)\right )^2 \mathrm{d}x\right\}
\]
where $f$ is the (unknown) density function. Since $f$ is unknown, the argmin of $\text{MISE}$ cannot be computed directly. Machine learning techniques or data driven heuristics can be applied. The two most common plug in methods are \textit{Scott's Method}~\cite{scott2015multivariate} and the \textit{Silverman Method}~\cite{silverman2018density} which are each built in to SciPy's \texttt{gaussian\_kde} method. For our use-case, the default Scott's method was sufficient; in this application $h$ is computed as:

\[
n^{-1/(d+4)}
\]
where $n$ is the number of points sampled and $d$ is the ambient dimension of the point set.

\section{Noise and Interference Calculations}

\subsection{SIR} \label{app:SIR}

To simulate realistic background interference, we create a background signal environment composed of multiple superimposed signals. The simulation uses the same ${5000 \text{ Hz}}$ sampling rate ($F_s$) as the comparison or main injection signal. This main signal of interest is explicitly defined also has a carrier frequency of ${100 \text{ Hz}}$, corresponding to its symbol rate of 100 symbols per second.
We model the interference using four distinct frequency bands (or channels) that are positioned around ${100 \text{ Hz}}$. The configuration is such that two bands are positioned below and two above the $100 \text{ Hz}$ main signal frequency. The channel bands are composed of five independent ${1 \text{ Hz}}$ interference sub-channels.
The sub-channels within a band are separated by a ${5 \text{ Hz}}$ frequency gap. 
The four main clusters are separated by an equal frequency gap of ${5 \text{ Hz}}$.
This results in a total of ${20}$ superimposed interference signals. The total occupied bandwidth of the four clusters and three gaps is 40 Hz, from 80 to 120 Hz.

The power of this simulated background interference is adjusted to achieve a specific Signal-to-Interference Ratio (SIR). This ensures the background interference has a controlled level of power relative to the primary signal, enabling us to test the system's performance under realistic interference conditions. To modify the amplitude of the background signal we sum them as $B(t) = \sum b(t)$ where $b(t)$ is a single sub-channel signal. This summing is based on what the observations of the background would be if filtered the entire RF spectrum to this bandwidth. The amplitude of $B(t)$ is then scaled according to the injection signals amplitude as
\begin{equation}
A_{B(t)} = \sqrt{\frac{P_{\text{injection}}}{10^{\frac{\text{SIR}{\text{dB}}}{10}}}}
\end{equation}
where $A{B(t)}$ is the amplitude of the combined background signal $B(t)$, $P_{\text{injection}}$ is the power of the injection signal, and $\text{SIR}_{\text{dB}}$ is the desired Signal-to-Interference Ratio in decibels.

\subsection{SNR} \label{app:SNR}
To determine the appropriate noise amplitude for a given signal and target Signal-to-Noise Ratio (SNR), we proceed as follows:
\begin{enumerate}
    \item \textbf{SNR Conversion:} We begin with the desired SNR expressed in decibels ($SNR_{dB}$) and convert it to a linear scale ($SNR$) using the following relationship:
    \begin{equation}
        SNR = 10^{\frac{SNR_{dB}}{10}}
    \end{equation}

    \item \textbf{Signal Power Calculation:}  We then calculate the power of the signal ($P_{signal}$) as the mean of the squared signal values:
    \begin{equation}
        P_{signal} = \frac{1}{N} \sum_{i=1}^{N} s[i]^2
    \end{equation}
    where $N$ represents the number of samples in the signal $s$.

    \item \textbf{Noise Power Determination:} With the linear SNR and signal power in hand, we calculate the required noise power ($P_{noise}$) using the definition of SNR:
    \begin{equation}
        SNR = \frac{P_{signal}}{P_{noise}} \implies P_{noise} = \frac{P_{signal}}{SNR}
    \end{equation}

    \item \textbf{Noise Amplitude (Standard Deviation):} Assuming additive white Gaussian noise (AWGN) with zero mean, the noise amplitude is represented by its standard deviation ($\sigma$). This standard deviation is related to the noise power by:
    \begin{equation}
        P_{noise} = \sigma^2 \implies \sigma = \sqrt{P_{noise}}
    \end{equation}
    This calculated $\sigma$ is then used as the standard deviation for generating the Gaussian noise samples.
\end{enumerate}

\section{Autoencoder Implementation}\label{app:AE}

A standard autoencoder neural network is implemented in this work to compress input data and subsequently reconstruct it for the purpose of anomaly detection. The core concept involves learning a lower-dimensional representation that captures the essential features of the input and measuring the quality of that representation in the networks ability to reconstruct the original vector. This architecture comprises two principal components: the encoder and the decoder.

The encoder's function is to reduce the dimensionality of the input. This is done by using a sequential network consisting of two linear layers, each with a ReLU activation applied after the first linear transformation. 
The output $\mathbf{z}$, set to 8 dimensions, represents the compressed version of $\mathbf{x}$.

The decoder reverses this process, attempting to reconstruct the original input from the compressed representation. It also consists of two linear layers, with a ReLU activation following the first, and a sigmoid activation at the output.
The output $\hat{\mathbf{x}}$ is the reconstructed input.

For training, the Mean Squared Error (MSE) loss function is used, which calculates the average squared difference between the input $\mathbf{x}$ and the reconstructed $\hat{\mathbf{x}}$:
\begin{equation}
{\rm Loss} = {\rm MSE}(\mathbf{x}, \hat{\mathbf{x}}) = \frac{1}{n} \sum_{i=1}^{n} (x_i - \hat{x}_i)^2,
\end{equation}
where $n$ is the dimension of the input vector. For our implementation we use an Adam optimizer and a learning rate of 0.001.

We use \texttt{PyTorch} for our autoencoder implementation. The data is processed in batches of 8, and the model is trained for 10 epochs. The autoencoder is applied using a sliding window approach: training on a baseline, followed by evaluation on subsequent data. This methodology allows for the detection of anomalies by observing increases in the reconstruction error when the evaluation data is not of the same form as the training baseline.
This process enables the identification of anomalies through changes in the reconstruction loss.

    

\end{document}